# On the Core Deuterium-Tritium Fuel Ratio and Temperature Measurements in DEMO

V.G. Kiptily

CCFE
Culham Science Centre, Abingdon, Oxon, X14 3DB
United Kingdom

E-mail: vasili.kiptily@ccfe.ac.uk

**Abstract**. Comparing with ITER, the experimental fusion machine under constraction, the next step test fusion power plant, DEMO will be characterized by very long pulse/steady-state operation and much higher plasma volume and fusion power. The substantially increased level of neutron and gamma fluxes will require reducing the physical access to the plant. It means some conventional diagnostics for the fusion plasma control will be not suitable in DEMO. Development of diagnostics along with the machine design is a primary task for the test plant. The deuterium-tritium fuel ratio and temperature are among important parameters, which should be under control. In this letter a novel technique for the core fuel ratio and temperature diagnostics is proposed. It is based on measurements and comparison of the rates $T(p,\gamma)^4He$ and $D(T,\gamma)^5He$ nuclear reactions that take place in the hot deuterium-tritium plasma. Based on detection of high energy gamma-rays, this diagnostic is robust, efficient and does not require direct access to the plasma. It could be included in the loop of the burning plasma control system. A feasibility of the diagnostic in experiments on JET and ITER is also discussed.

A goal of the DEMO fusion power plant is to demonstrate that the major scientific and technological obstacles on the way to the commercial power plant are overcome. And one of the real challenges is development of the unprecedented robust and reliable diagnostic system [1, 2]. A harsh DEMO environment with a very high level of neutron and gamma-ray fluxes will make some conventional ITER diagnostics unfeasible. Among the restricted set of instruments, which are available for the machine protection and plasma control in DEMO, neutrons and gamma-ray measurements will be useful as the neutron and gamma detectors can be placed far away from the plasma and they do not require a direct access to the vessel.

The main roles of neutron diagnostics in DEMO will be fusion power and core ion temperature measurements [3, 4]. Gamma-ray diagnostics [5], which are routinely used for fast ion studies on JET [6], can provide information on the confined alpha-particle distribution in the energy range $E_\alpha$>1.7 MeV, impurities and fusion power in ITER. In this letter, a novel gamma-ray technique is proposed, which can allow measurement of the core fuel ratio in DEMO.

A conventional diagnostic for the direct measurement of the plasma composition is a neutral particle analyser (NPA). A prototype of the NPA system developed for the ITER operation [7] has been tested and successfully used on JET [8]. A feasibility study of the method for the fuel ratio measurements in DEMO is still required. The NPA detectors should be placed far away from the reactor and shielded from neutrons. In DEMO, the detected neutral particles will be strongly weighted to those from the plasma periphery, where the neutral gas pressure is highest, and it may be difficult to de-convolve the spectra to yield the core fuel composition reliably.

Neutron spectrometry is another tool for the fuel ratio measurement [9]. Indeed, the ratio of neutron rates generated in the $D(D,n)^3He$ and $D(T,n)^4He$ reactions is proportional to the fuel ratio, $n_D/n_T$ (DTR). For the steady state plasma this relation is rather simple

$$\frac{R_{DDn}}{R_{DTn}} = \frac{1}{2}\frac{n_D}{n_T}\frac{\langle\sigma v\rangle_{DD}}{\langle\sigma v\rangle_{DT}},$$

where $\langle\sigma v\rangle_{DD}$ and $\langle\sigma v\rangle_{DT}$ are reactivities in the case of the Maxwellian distribution function of the plasma components (Fig.1). The neutron spectrum of DT plasmas consists of several main components: two peaks, one at 2.5 MeV due to the DD-reaction and another at 14 MeV due to the DT-reaction, and three components with continuous neutron spectra, at $E_n < 11.5$ MeV from the three-body reaction $T(T,2n)^4He$, at $E_n < 2.5$ MeV from endothermic secondary reaction $T(p,n)^3He$ and a continuous spectrum of the scattered neutrons. Cross-sections of the above mentioned reactions are shown in Fig.2.

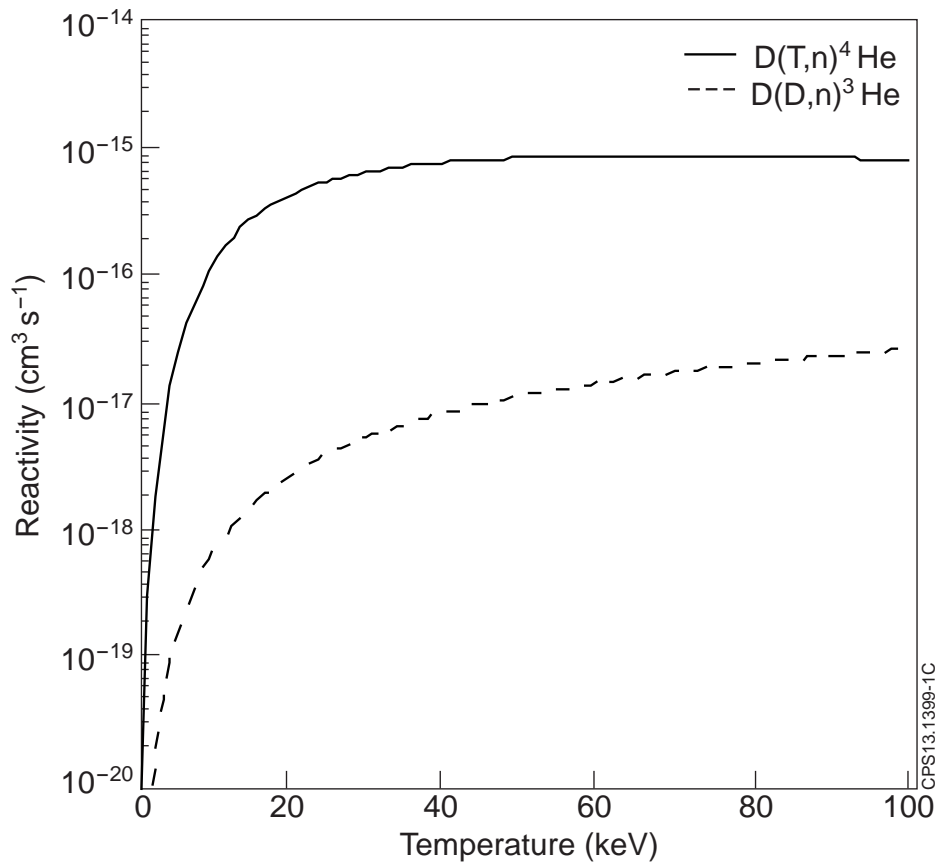

**Figure 1.** Reactivities of the $D(D,n)^3He$ and $D(T,n)^4He$ reactions calculated with parameters for Maxwellian plasmas from [10].

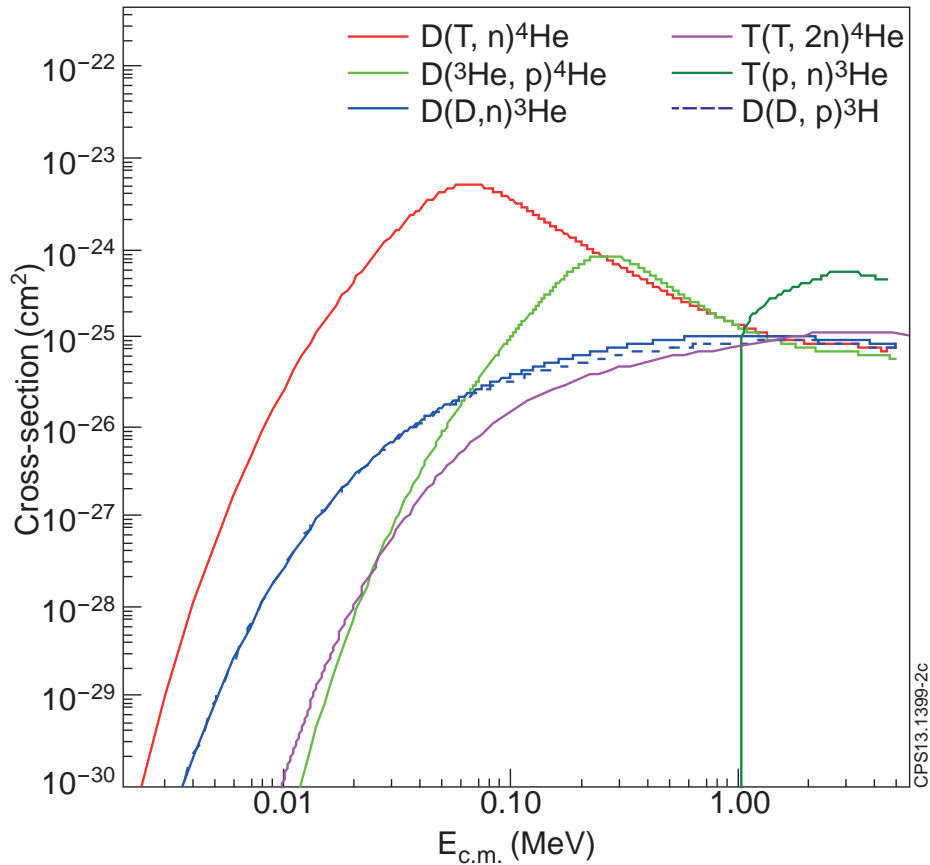

**Figure 2.** Cross-sections of main fusion reactions calculated with parameters from Ref.10 in comparison with *T(T,2n)⁴He* [11] and *T(p,n)³He* [12].

A main applicability criterion of this diagnostic approach is the signal-to-background ratio (*S/B*) for the 2.5-MeV peak. In the feasibility study of the diagnostic proposed for the ITER application [13] has been found that the main feasibility concern is scattered neutrons. It was shown that for the *DTR*=1 the total *S/B* ratio for the 2.5-MeV peak is around of 0.4 in the case of the tangential line-of-sight. The radial plasma observation gives much worse *S/B* value. Neutrons from TT-reaction play a role in the case of $n_D/n_T \equiv DTR < 1$. However, influence of the continuous neutron spectra from the secondary reactions of bulk tritium with fusion protons and tritons, for example *T(p,n)³He* reaction need to be assessed.

The gamma-ray diagnostics has some significant advantages to be used on DEMO. First of all, gamma-ray spectrometry of the plasma does not require a direct access to the vessel. Furthermore, several centimetres of the first wall could play a positive role as it reduces the strong flux of the undesirable low energy gamma-rays. Gamma-ray detectors are compact and the response function is well known and simple. Viewing the plasma through a long collimator, the spectrometer can be placed far away from the machine. The collimator should be filled with material, which suppress the neutron flux and thus neutron induced gamma-ray background. In this case, the required radiation conditions behind the biological shield of DEMO will be assured. Gamma-ray detectors can be easily replaced in the case of any damage.

There are several nuclear reactions that could be useful for the fuel ratio measurements. List of these reactions $A(B,\gamma)C$ is given in the Table 1. The nuclear reaction energies, the *Q*-values, which characterise the mass balance of the reactions $Q = (M_A + M_B - M_C)c^2$, are also presented there and can be used for an assessment of the excitation energy of the residual nuclei. In the centre-of-mass frame, the energy of gamma-rays de-excited the final nuclei to the ground state is given by $E_\gamma = Q + E_A + E_B - E_C$. For some reactions a ratio of the radiation capture to the cross-section of a main fusion reaction (branching ratio or BR) is available and presented in the Table 1.

One can see, by analogy with the above mentioned neutron diagnostic, $D(D,\gamma)^4He$ and $D(T,\gamma)^5He$ radiation capture reactions could be useful for the fuel ratio diagnostic, but the cross-section of the $D(D,\gamma)^4He$ reaction is very low as the branching ratio of two orders of magnitude less than the $D(T,\gamma)^5He$.

There is another group of the radiation capture reactions, which are suitable for this purpose. Indeed, secondary reactions of fuel with energetic charged products from the fusion reactions $D + D \rightarrow ^3He(0.82 MeV) + n(2.45 MeV)$ and $D + D \rightarrow t(1.01 MeV) + p(3.02 MeV)$ can be used. The rate of these fusion reactions is proportional to $n_D^2$, and secondary reactions $D(p,\gamma)^3He$, $D(t,\gamma)^5He$, $D(^3He,\gamma)^5Li$ and $T(p,\gamma)^4He$ are also proportional to $n_D$ and $n_T$. However, not all of these reactions are feasible for the fuel ratio measurements.

The first reaction in the Table 1 has relatively large cross-section and has been exploited for the effective temperature assessments of hydrogen ions injected for the plasma heating (NBI) [18] and accelerated during the minority ion cyclotron resonance heating (ICRH) [6]. In the case of Maxwellian distribution function of the interacted ions, the gamma-ray peak position $E_\gamma = Q + E_G$, depends on the Gamov's peak energy, for example for the $D(p,\gamma)^3He$ reaction $E_G \approx 0.74 \langle T_{Dp} \rangle^{2/3}$ (MeV). The peak width also depends on $E_G$ and the Doppler broadening due to the reaction kinematics. The experiments with the H-minority ICRH have shown that this diagnostic tool is useful for the ion tail temperatures below 400 keV. The broadening at high temperatures makes the peak unrecognisable since the background, which continues up to 9 MeV. In the DEMO case, the fusion protons, with a broad energy distribution below 3 MeV, will give rise to excessively broad spectrum of gammas from the $D(p,\gamma)^3He$ reaction and intensity of the peak cannot be convincingly obtained.

**Table 1.** Summary of the radiation capture reactions potential for the fuel ratio measurements

| Reaction | Q, MeV | a) Branching ratio |
|---|---|---|
| $D(p,\gamma)^3He$ | 5.5 | - |
| $D(D,\gamma)^4He$ b) | 23.84 | $10^{-7} - 10^{-6}$ |
| $D(T,\gamma)^5He$ c) | 16.63 | $5 \times 10^{-5} - 5 \times 10^{-4}$ |
| $D(^3He,\gamma)^5Li$ d) | 16.38 | $5 \times 10^{-5} - 5 \times 10^{-4}$ e) |
| $T(p,\gamma)^4He$ f) | 19.814 | - |

a) In the energy range 0.02 – 3 MeV.
b), c), d) Branch reactions: $D(D,n)^3He$ and $D(D,p)^3H$ [14], $D(T,n)^4He$ [15, 16], $D(^3He,p)^4He$ [17]; e) for the ground state branch.
f) Branch reaction - $T(p,n)^3He$ with $Q$ = -0.764 MeV (threshold reaction)

The $D(^3He,\gamma)^5Li$ reaction has been used for the assessments of the fusion power generated in the plasma with $D(^3He,p)^4He$ reaction [19] and the efficiency of the $^3He$ minority ICRF heating in $D^3He$ plasmas [6], but it is inappropriate for the for the DT fuel ratio measurements in DEMO. There is an advantage that gamma-ray energy spectrum generated by $^3He$ ions lies outside of the energy range with a strong background, however spectra of the $D(^3He,\gamma)^5Li$ and $D(T,\gamma)^5He$ reactions are overlapped as Q values are roughly the same. Furthermore, because of the large width of the $^5Li$ ground state ($\Gamma_{g.s.}$=2.6 ± 0.4 MeV) and the first excited one ($E_x$ = 7.5 ± 1.0 MeV, $\Gamma_x$=6.6 ± 1.2 MeV), the peaks cannot be separated. Although both $D(^3He,p)^4He$ and $D(T,n)^4He$ reaction cross-sections are nearly the same in the range above 0.2 MeV (Fig.2), the yield of the secondary reactions is much lower and the $D(^3He,\gamma)^5Li$ gamma-rays will be a background for the intensive gamma-ray peak from the $D(T,\gamma)^5He$ reaction.

To measure the fuel ratio in DEMO with the gamma-ray diagnostic, utilizing of the $T(p,\gamma)^4He$ and $D(T,\gamma)^5He$ reactions would appear to be the only practical possibility, deriving the $n_D/n_T$ value from the ratio of gamma-ray reaction rates $R_{Tp\gamma}/R_{DT\gamma}$. The $T(p,\gamma)^4He$ reaction has been proposed for using in ITER [20].

Indeed, the gamma-ray peaks related to these reactions are well separated and both lies far away from the strong background energy range. The main gamma-ray background is build up by the neutron capture $(n,\gamma)$ and inelastic scattering $(n,n'\gamma)$ reactions in the range $E_\gamma < 9$ MeV. In the range $E_\gamma < 14$ MeV background mainly produced in the $(n,n'\gamma)$ reactions. However, density of nuclear levels at high excitation energies is very low that is why background gamma-ray emission in the range 9-14 MeV is rather weak. Gamma-rays with energies $E_\gamma > 14$ MeV can be produced in radiation capture nuclear reactions between the possible low-Z plasma impurities and confined charged particles in the plasma, e.g. $^7Li(p,\gamma)^8Be$ (Q=17.35MeV), $^{11}B(d,\gamma)^8Be$ (Q=18.7MeV), $^{14}N(d,\gamma)^{16}O$ (Q=20.7 MeV). However, both concentration of impurities in plasmas and cross-sections of these reactions are very low. The $D(^3He,\gamma)^5Li$ gamma-rays which are a weak background for the $D(T,\gamma)^5He$ can be taken into account for the fuel composition measurements.

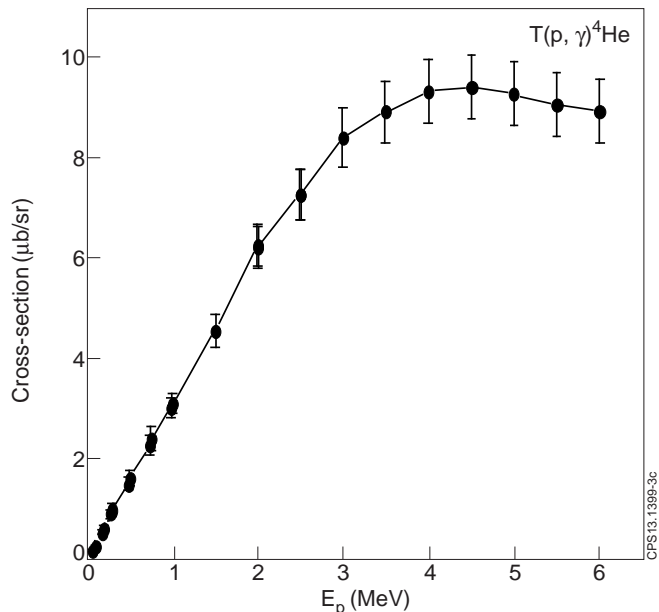

**Figure 3.** The experimental differential cross-section of the $T(p,\gamma)^4He$ reaction [21,22].

Experimental studies of the $T(p,\gamma)^4He$ reaction [21, 22] shows that its cross-section (Fig.3) is high enough and exceeds the cross-section of the $D(^3He,\gamma)^5Li$ reaction by factor of 10 in the MeV range. It has been demonstrated that the $T(p,\gamma)^4He$ reaction can be used for characterisation of the fast H-ion tail during the ICRF heating of tritium plasmas in JET [23] as the produced gamma-ray spectrum depends on the proton distribution function. The Compton part of the $T(p,\gamma)^4He$ gamma-ray spectrum was already recorded in the JET experiments with fully tritium plasmas during H-minority heating [24].

For the feasibility assessment of the fuel ratio measurements in the DEMO plasma core by means of the gamma-ray diagnostic, two possible reactor designs: one for the steady state (SS) operation, with the major radius $R$=8.5m and the minor radius $a$=2.83m and one for the long pulse operation (LP) with $R$=9.6m and $a$=2.4m , are used [25]. Parameters for SS and LP operations considered in this paper with two values for plasma current, central electron density and central electron temperature related to the peaked density and flat density scenarios [25] presented in the Table 2 and used for a consistent calculations sensitivity of the method. The effective charges $Z_{eff}$=2.57 and 1.95 for SS and LP operations, correspondingly, were used for the fuel density calculations, assuming the average impurity charge $Z_{imp}$=6. The parameter $Z_{imp}$ is important one for the fusion reactor design as it defines the fuel density and by that the fusion power. Indeed, the fuel density $n_{DT}$ depends on both $Z_{eff}$ and $Z_{imp}$ as follow

**Table 2.** DEMO parameters [25] used for the feasibility study of the fuel ratio diagnostic.

| DEMO plasma scenarios | | $n_{e0}$ ($10^{19}$ m$^{-3}$) | $T_{e0}$ (keV) |
|---|---|---|---|
| Steady-state | $n_e$-flat | 9.3 | 64 |
| | $n_e$-peaked | 15.0 | 53 |
| Long Pulse | $n_e$-flat | 10.4 | 54 |
| | $n_e$-peaked | 16.8 | 57 |

$$\frac{n_{DT}}{n_e} = \frac{Z_{imp} - Z_{eff}}{Z_{imp} - 1}.$$

One can see that increase of $Z_{imp}$ lead to increase of $n_{DT}$ and, consequently fusion reaction rates. In this paper, the chosen value $Z_{imp}$ is consistent with the expected DEMO plasma content though there is no an established model yet. For the SS case with $Z_{eff}$=2.57, assuming that main impurities are *He*, *Ar* and *W* and their relative concentrations 8%, 0.3% and 0.0005%, the parameter $Z_{imp}$≈6.2. If the plasma contains *He*, *Li*, *Ar* and *W* with relative concentrations 6%, 2%, 0.28% and 0.0005%, the parameter will be around of 5.9. In this case, rates of the $D(T,\gamma)^5He$ and $T(p,\gamma)^4He$ reactions will be 5% and 7% less, correspondingly. Here, lithium is a proxy for some seeding by as yet unspecified low atomic number impurities to control radiated power fraction, and not critical to the main topic of the paper.

Studying sensitivity of the gamma-ray diagnostic to the $n_D/n_T$ value, the reaction rate ratio of the $T(p,\gamma)^4He$ and $D(T,\gamma)^5He$ reactions was examined for different steady Maxwellian plasmas. On the Fig.4 rates of these reactions calculated for the SS scenario with flat $n_e$-profile (see Table 2) are presented against the $n_D/n_T$ value. It was assumed that $n_D$-, $n_T$- and $Z_{eff}$ -profiles have the same shape as $n_e$-profile.

For the $T(p,\gamma)^4He$ reaction rate the DD fusion proton energy distribution function was calculated in the simple classical approximation [26], i.e.

$$f_p(E) \propto \frac{1}{E_0}\sqrt{\frac{E_0}{E}}\frac{E_0^{\frac{3}{2}} + E_c^{\frac{3}{2}}}{E^{\frac{3}{2}} + E_c^{\frac{3}{2}}},$$

with the critical proton energy $E_c \propto T_e$ and the Spitzer slowing-down time $\tau_e \propto T_e^{3/2}/n_e$. Since the distribution function depends on temperature, gamma-rays of the $T(p,\gamma)^4He$ reaction could be used to infer this important plasma parameter. In reality, the proton energy distribution function in DEMO can differ from this one, especially in the energy range below the critical energy at around of 0.9 MeV, where the transport effects and losses may play a role. However, the proton energy distribution function can be derived from the $T(p,\gamma)^4He$ gamma-ray spectrum recorded with a contemporary scintillation spectrometer. An example of how the spectrum depends on the $T_e$ with the classical slowing down demonstrated in Fig.5a.

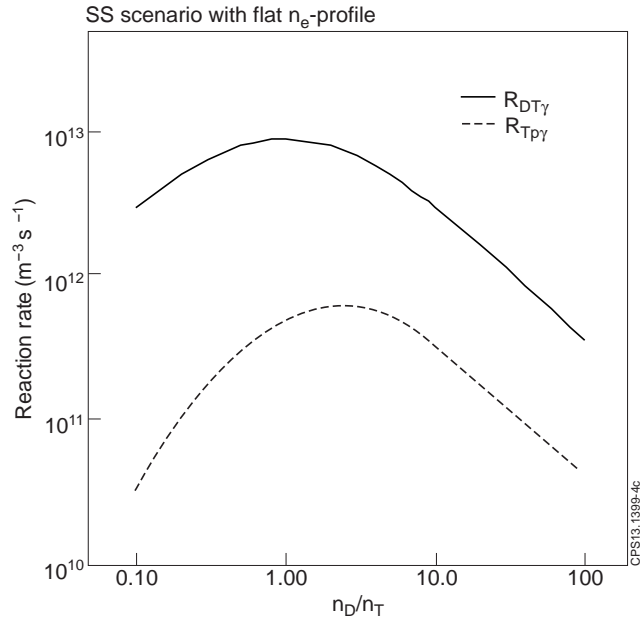

**Figure 4.** Rates of the $T(p,\gamma)^4He$ and $D(T,\gamma)^5He$ reactions vs. $n_D/n_T$ calculated for the SS scenario with flat $n_e$-profile (see Table 2).

The temperature dependence of the reaction rate is shown in Fig.5b. With increasing of the electron temperature, the maximum of the $T(p,\gamma)^4He$ reaction emission is shifted to the higher energy at about of 1 MeV and also the peak became much broader. These changes are due to the changes of the fusion protons energy distribution and which are big enough for obtaining information about $<T_e>$ averaged over period, which could provide required accuracy for the de-convolution of the continuously recorded $T(p,\gamma)^4He$ spectrum. It is important to emphasise that the $n_e$-dependence of the gamma-ray spectrum is negligible. So, in DEMO the $<T_e>$ could be monitored in the real time mode with a resolution time defined by the $T(p,\gamma)^4He$ gamma-ray count-rate. For this task a comprehensive proton distribution function is needed, which should account the protons transport, alpha-particle knock effects etc. A special study of the issue should be addressed to ITER experiments for a validation of this model.

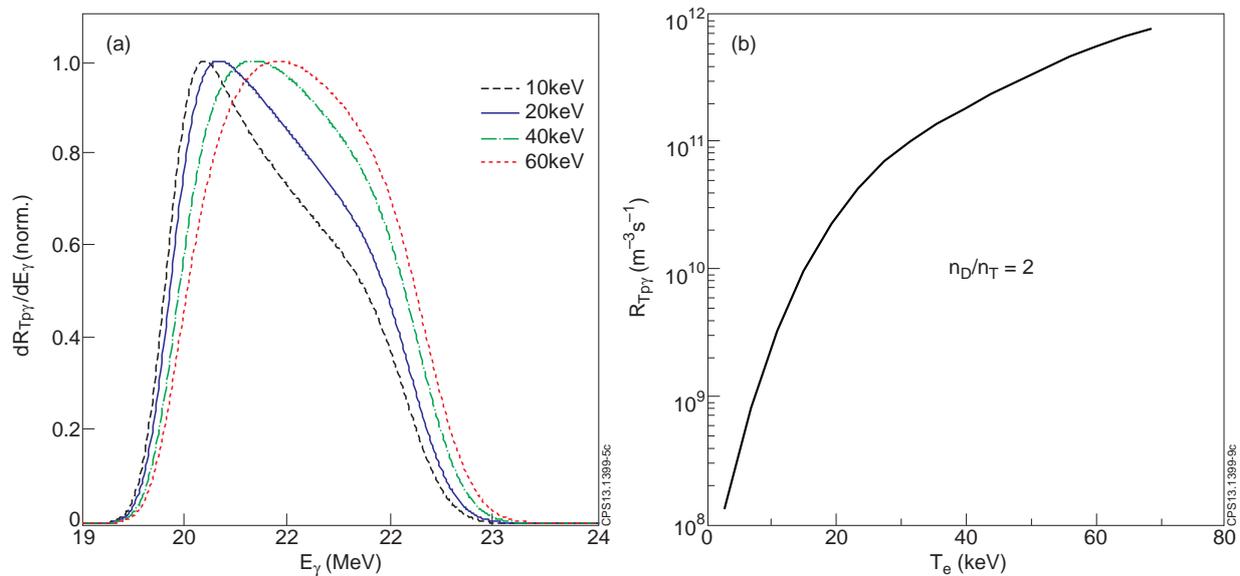

**Figure 5.** (a) – $T(p,\gamma)^4He$ gamma-ray spectra calculated at different $T_e(0)$; (b) – $T(p,\gamma)^4He$ reaction rate vs. $T_e(0)$ for the case $n_D/n_T=2$. The calculations have been done for the SS scenario with flat $n_e$-profile (see Table 2); Te-profiles were assumed the same.

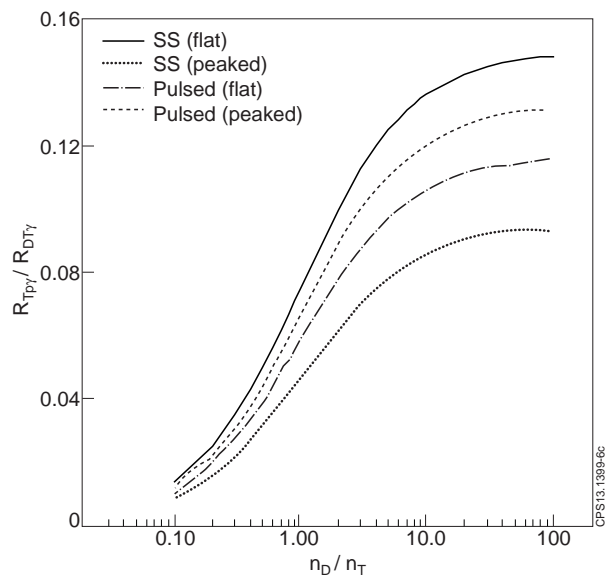

**Figure 6.** Ratio of the $T(p,\gamma)^4He$ and $D(T,\gamma)^5He$ reaction rates (like on Fig.4) vs. $n_D / n_T$ calculated for the DEMO scenarios (see Table 2) used for the feasibility study.

The $R_{T p\gamma}/R_{DT\gamma}$ ratios calculated for all DEMO scenarios presented in Table 2 are shown in Fig. 6. One can see that sensitivity of this diagnostic lies in the $n_D/n_T$ range of 0.1 – 10, with a particularly strong sensitivity around the expected reactor operating point at or near $n_D/n_T$ =1.0. The temperature dependence of the ratio and the relevant spectrum measurement uncertainties shown in Fig.7 demonstrates an impact of temperature changes and the sensitivity of the proposed technique. However, to assess the experimental accuracy of the fuel ratio measurements, a detailed scheme of the diagnostic should be considered.

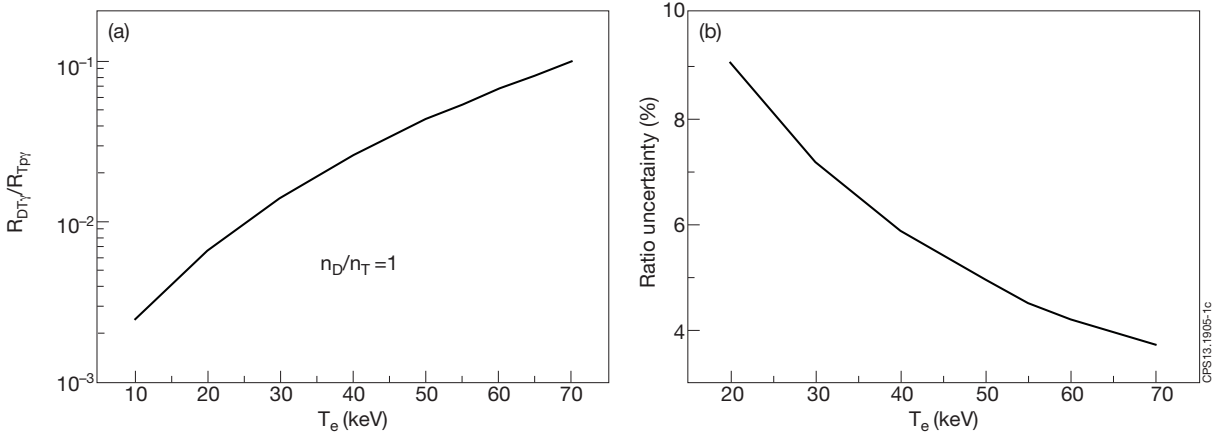

**Figure 7.** (a) – rate ratio for $T(p,\gamma)^4He$ and $D(T,\gamma)^5He$ gammas calculated vs. $T_e(0)$ in the case $n_D/n_T$=1; (b) – uncertainties of the reaction ratio vs. $T_e(0)$. The calculations have been done for the SS scenario with flat $n_e$-profile (see Table 2); Te-profiles were assumed the same.

For the proposed gamma-ray measurements in DEMO a well-collimated spectrometer is needed. It should be fast, providing a MHz count-rate and its energy resolution should be sufficient to give the proton energy distribution reconstruction with a required accuracy. To illustrate the application of this technique as DEMO diagnostics, and assess the diagnostic time resolution and measurement accuracy, it is assumed that the spectrometer is placed in 20 m from the plasma centre and it has tangential line-of-sight viewing the plasma through the 2-m long collimator. The collimator with diameter of 4-cm is filled up by a material, which can effectively attenuate neutrons and has a high transparency for MeV gamma-rays. The best material is lithium hydride (*LiH*), which is already used on JET [27]. A 30-cm sample of the $^6LiH$-filter reduces 2.5-MeV neutron flux by factor of 900 and the 15-MeV neutron flux by factor of 30 [28]. The 2-m long collimator plugged with *LiH* will attenuate 14-MeV neutrons by factor of $2 \times 10^{10}$ and gamma-rays in the energy range 15 – 25 MeV by factor of 100.

To provide high detection efficiency for the $T(p,\gamma)^4He$ and $D(T,\gamma)^5He$ gammas, a full energy peak spectrometer [5] should be used with efficiency up to 60%. Nowadays the best detector for this diagnostic is a heavy high-Z scintillator *LaBr$_3$(Ce)*. It has short decay time (~20 ns), high photons yield and low sensitivity to neutrons. The high rate capability could be enabled by a dedicated data acquisition system with a sampling frequency of 400 MHz or higher and 14-bit resolution [29]. These outstanding properties of the system open a possibility to extend the pulse height analysis limit beyond of 5 MHz [30] with energy resolution ~1% in the energy range 15 – 25 MeV.

Using this diagnostic scheme, the $T(p,\gamma)^4He$ gamma-ray count rate will depend on the $n_D/n_T$ ratio as shown in Fig.8. These results reveal that proposed method is feasible for measurements of fuel ratio in the range 0.5 – 10.0. Note that in this work the DD fusion source of fast protons is taken into account only. There are some other channels to increase the rate of this reaction, e.g. interaction of DD fusion tritons with bulk hydrogen, $H(t,\gamma)^4He$. However, relative $H$ concentration will be low; hence this additional contribution to the reaction rate is small. Also, an alpha-particle knock-on effect could increase rates of the $D(d,p)T$ and $D(t,n)^4He$ reactions, enhancing gamma-ray yields of $T(p,\gamma)^4He$ and $H(t,\gamma)^4He$ and $D(t,\gamma)^5He$ reactions. In the JET deuterium tritium experiments have been found the knock-on affect on the tail of the neutron spectrum [31], but the relative contribution these neutrons is rather small. Indeed, during the implementation of the diagnostic for the core fuel ration control, all useful reactions and effects should be taken into account for the modelling of the complete proton distribution function. In this Letter a simplified approach is used for the assessment of the technique feasibility.

Since the $D(T,\gamma)^5He$ gamma-ray rate will be higher by factor of 7 – 70 in the $n_D/n_T$ value range, the main contribution to the statistical uncertainty of this diagnostic will give measurement of the $T(p,\gamma)^4He$ gamma-rays. Figure 9 demonstrates the relative uncertainties of measurements for the acquisition time t=1.0 s. In the DEMO with the steady state operation and slowing-down time $\tau_e \sim$ 5-7s, the fuel composition variations in the core is expected to be rather slow. In this case, the accuracy $\varepsilon$ of the measurements could be improved by means of increasing the exposure time, so that $\varepsilon \propto 1/\sqrt{t}$. A continued recording of the $T(p,\gamma)^4He$ and $D(T,\gamma)^5He$ gamma-ray spectra will allow monitoring of the fuel ratio and making corrections of the fusion proton energy distribution function derived from the shape of the $T(p,\gamma)^4He$ spectra.

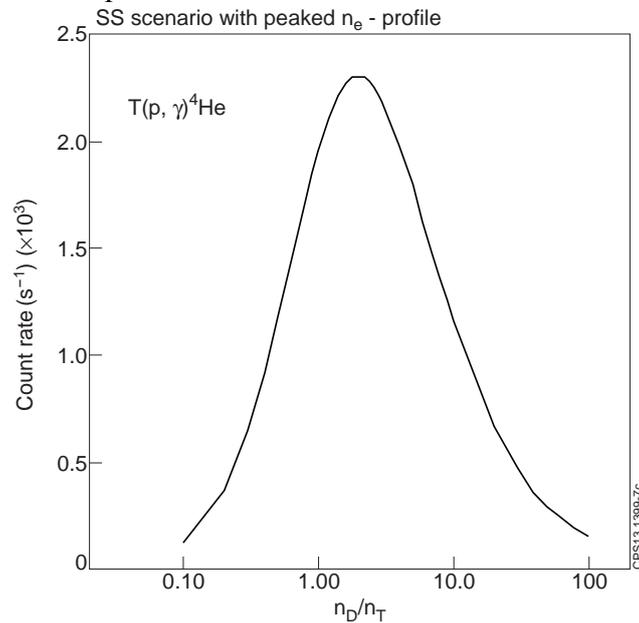

**Figure 8.** Count rate of the $T(p,\gamma)^4He$ gammas vs. $n_D/n_T$ calculated for the SS scenario with peaked $n_e$-profile (see Table 2).

The accuracy presented in Fig.9 reflects expected experimental uncertainties of the $R_{Tp\gamma}/R_{DT\gamma}$ ratio measurements. The uncertainty of the $n_D/n_T$ ratio inference will be higher as it will include systematic uncertainties connected to reconstruction of the fusion protons distribution function (the model is still an issue), nuclear reaction cross-sections, spectrometer calibration etc. Note that measurements of relative changes of the fuel ratio in core for the continuous plasma control could reduce systematic errors.

As was shown in [23], the proposed technique can be tested on JET during H-minority heating of the tritium plasma, where the effective tail temperature of accelerated H-ions can be inferred form the $T(p,\gamma)^4He$ spectra. For the application of the proposed fuel composition measurements in the high performance DT discharges, a well-shielded gamma-ray spectrometer with the high count rate capabilities [30] is needed.

According to the ITER requirements on the fuel ratio measurements in core, accuracy of 20% should be provided during of 100 ms acquisition time. For the ITER steady state scenario [32] with $R$=6.35m, $a$=1.85m, $n_e$=8.5x10$^{19}$m$^{-3}$, $Z_{eff}$=2.6, $T_e$=25keV and $n_D/n_T$=1, the required accuracy will be obtained recording gamma-ray spectrum during of 150 ms. For the assessments the diagnostic setup was taken the same as in the DEMO case, but the length of the *LiH*-plug in the collimator has been reduced by factor of 2. This diagnostic will be feasible in ITER, but an optimisation of the gamma-ray spectrometry system is needed to meet the requirements.

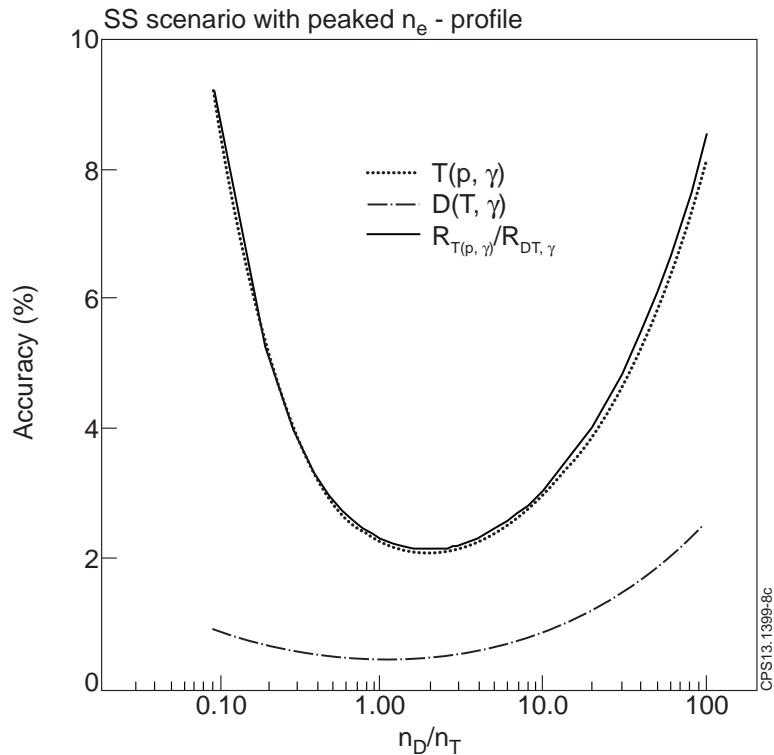

**Figure 9.** Statistical uncertainty of the *T(p,γ)$^4$He* and *D(T,γ)$^5$He* gamma-ray measurements and its ratio vs. $n_D/n_T$ calculated for the acquisition time of 1s in the SS scenario with peaked n$_e$-profile.

Concluding the paper, one can say that JET, together with ITER experiments, operating with DT plasmas will provide the opportunity to obtain full information on the feasibility and capability of the proposed technique for the temperature measurements and control of the fuel composition in the DEMO plasma core with time resolution smaller than the energy and particle confinement time.

**Acknowledgments**

Credit and thanks are due to Tom Todd, William Morris, Sergei Sharapov, Ken McClements and Victor Yavorskij for fruitful discussions of the work.

This work was funded by the RCUK Energy Programme [grant number EP/I501045] and the European Communities under the contract of Association between EURATOM and CCFE. To obtain further information on the data and models underlying this paper please contact PublicationsManager@ccfe.ac.uk . The views and opinions expressed herein do not necessarily reflect those of the European Commission.